\documentclass{ptapap}

\author{Micha{\l} J.~Micha{\l}owski}[ROE]
\affil[ROE]{SUPA (Scottish Universities Physics Alliance), Institute for Astronomy, University of Edinburgh, Royal Observatory, Edinburgh, EH9 3HJ, UK}

\title{Dust production $0.7$--$1.5$ billion years after the Big Bang}

\newcommand{\msun}{\mbox{$M_\odot$}}
\newcommand{\submm}{submillimetre}
\newcommand{\mstar}{M_{\rm star}}

\begin{document}

\maketitle

\begin{abstract}

Cosmic dust is an important component of the Universe, and its origin, especially at high redshifts, is still unknown. I present a simple but powerful method of assessing whether dust observed in a given galaxy could in principle have been formed by asymptotic giant branch (AGB) stars or supernovae (SNe). Using this method I show that for most of the galaxies with detected dust emission between $z=4$ and $z=7.5$ ($1.5$--$0.7$ billion years after the Big Bang) AGB stars are not numerous and efficient enough to be responsible for the measured dust masses. Supernovae could account for most of the dust, but only if all of them had efficiencies close to the maximal theoretically allowed value. This suggests that a different mechanism is responsible for dust production at high redshifts, and the most likely possibility is the grain growth in the interstellar medium.

\end{abstract}

\section{Introduction}

Cosmic dust is an important component of the Universe, because dust can be found in almost every galaxy, and because half of the energy ever emitted by stars has been absorbed by dust, as shown by the cosmic optical and infrared background \citep[e.g.][]{hauserdwek01,dole06}. Dust is known to be present at all cosmic epochs, and has been detected at redshifts as high as $z\sim7.5$ \citep{watson15}. 

The outstanding question is how dust is formed, and there are three most accepted options, which may all contribute to dust production depending on a galaxy's type and redshift. Dust is known to be produced in the atmospheres of asymptotic giant branch (AGB) stars \citep{meixner06,matsuura09,matsuura13,sloan09,srinivasan09,boyer11,boyer12,riebel12}, and theoretical works showed that the AGB dust production depends on a star's mass and metallicity, reaching the maximum value of  $\sim0.04\,\msun$ of dust per star \citep{morgan03,ferrarotti06,ventura12,nanni13,nanni14,schneider14}. On the other hand some supernovae (SNe) were observed to produce a significant amounts of dust \citep[$\sim1\,\msun$;][]{dunne03,dunne09casA,morgan03b,gomez09,matsuura11,indebetouw14,gomez12,temim13}, consistently with theoretical models \citep{todini01,nozawa03}. However, such `maximum' SN dust production has been claimed to be rare \citep{gallagher12,temim12}. Finally, it is possible that stellar sources only form small amount of dust, and the bulk of the dust mass accumulation happens thorough grain growth in the interstellar medium \citep[ISM;][]{draine79,dwek80,draine90,draine09}.
 
In the Milky Way and local galaxies most of the stardust (dust formed directly by stellar sources) has been attributed to AGB stars  \citep{gehrz89,zhukovska13}. The same has also been claimed for some high-redshift quasars \citep{valiante09,valiante11}, but this is more controversial. On the other hand, based on the comparison of the measured dust masses with estimated number of dust-producing stars, most other works lean towards SNe as the most efficient dust producers, but with a significant contribution of the grain growth in the ISM \citep{dwek07, dwek11,dwek11b,dwek14,dwek15,michalowski10smg4,michalowski10qso,michalowski15,gall11b,hjorth14,rowlands14b,zavala15}.
 Direct evidence of SN-synthesised dust is scarce, as this was obtained only for three objects, based on their flat extinction curves: $z\sim6.2$ quasar (\citealt{maiolino04,gallerani10}; but see \citealt{hjorth13}) and two GRB host galaxies at $z\sim6.3$ (\citealt{stratta07b}; but see \citealt{zafar10,zafar11}) and  $z\sim5$ \citep{perley10}.

In this paper I  present the method I developed to address the issue of dust production, and  review the results obtained for galaxies between $z=4$ and $z=7.5$ ($1.5$--$0.7$ billion years after the Big Bang). 

\section{Sample selection}

I discuss here the results based on galaxies at $z>4$ for which dust emission has been detected. This includes \submm-selected galaxies at $4<z<5$ \citep[discussed in][]{michalowski10smg4}, quasars at $5<z<6.5$  \citep[discussed in][]{michalowski10qso} and  dust-detected galaxies at $z>6.3$: {\it Herschel} selected galaxy at $z\sim6.34$ \citep{riechers13}, a quasar at $z\sim7.1$ \citep{mortlock11,venemans12} and a Lyman break galaxy at $z\sim7.5$ \citep{watson15}.

Moreover, in order to obtain information on upper limits of the dust yield per star \citep[see also][]{hirashita14}, I also analysed galaxies at $6.3<z<7.5$ for which dust emission has not been detected 
\citep{hu02,kanekar13,bradley12,schaerer14,ouchi09,ouchi13,iye06,ota14,finkelstein13}.

\section{Method}

Dust masses for all galaxies were estimated with a spectral energy distribution (SED) modelling, either by a panchromatic treatment including all wavelengths, or by a grey-body fits, which usually required the dust temperature and emissivitiy index to be assumed. 

Stellar masses of  galaxies whose emission is dominated by star-formation were calculated from the SED modelling \citep[the method described][]{michalowski08,michalowski09,michalowski10smg}. The assumption of double-component star formation histories resulted in higher stellar masses \citep{michalowski12mass,michalowski14mass}, which led to more conservative results (more potential dust-producing stars).

For quasars optical/near-infrared emission contains no information about stellar populations, so stellar masses were obtained from dynamical ($M_{\rm dyn}$) and gas ($M_{\rm gas}$) masses measured from {\submm} lines (carbon monoxide or atomic carbon). Stellar masses were estimated as $M_{\rm star}=M_{\rm dyn}-M_{\rm gas}$, or at least an upper limit could be set as $M_{\rm star}<M_{\rm dyn}$.

In a given galaxy the number of stars with masses between $M_0$ and $M_1$ in a stellar population with a range of masses between  $M_{\rm min}$ and $M_{\rm max}$ can be calculated from stellar mass of this galaxy and the slope of the initial mass function (IMF; $\alpha$):
\begin{equation}   
N(M_0-M_1)=\mstar \int_{M_0}^{M_1} M^{-\alpha} dM / \int_{M_{\rm min}}^{M_{\rm max}} M^{-\alpha}  M dM, 
\end{equation}
where  $(M_0,M_1)=(1,8)\,\msun$ for AGB stars and $(8,40)\,\msun$ for SNe, respectively. The lower limit for AGB stars may be adjusted to take into account only stars which had finished their main-sequence life at a given redshift (and had started producing dust), assuming that they were born shortly after the Big Bang. Hence, it was assumed to be $2.5\,\msun$ at $4<z<6$, and  $3\,\msun$ at $z>6$.

Finally, the average dust yield per  star required to explain the observed dust masses is $M_{\rm dust} /  N(M_0$--$M_1)$. This dust yield can be expressed as  $f\times(M_{\rm dust}/\mstar)$, where $f=\int_{M_{\rm min}}^{M_{\rm max}} M^{-\alpha}  M dM / \int_{M_0}^{M_1} M^{-\alpha} dM$. The compilation of the values of the factor $f$ for various types of stars and IMFs can be found in Table~2 of \citet{michalowski15}. The derived required dust yield can be compared with the theoretically allowed values to assess whether a given type of stars could be responsible for dust production in a given galaxy.

\section{Results and Discussion}

The analysis of {\submm} galaxies at $4<z<5$ \citep{michalowski10smg4} resulted in the required yields per AGB star of $\sim0.1\,\msun$ \citep[Fig.~1 of][]{michalowski11}. This is significantly above the theoretically allowed value, which rules out significant contribution of AGB stars to dust production in these objects. The required yields for SNe are $\sim0.2$--$0.7\,\msun$. Hence it is possible that SNe were responsible for dust production in these galaxies, but almost all SNe would need to be maximally efficient to account for the observed dust mass.

For $5<z<6.5$ quasars the constraints are even stronger. The required dust yields per AGB star are a few times  $0.1\,\msun$, and per SN are close to or exceeding $1\,\msun$ \citep[Fig.~1 of][]{michalowski10qso}. We also demonstrated that even with the combined effort of AGB stars and SNe, it is difficult to explain the observed dust masses.

For the $z\sim7.1$ quasar the situation is less conclusive as the required yields  extend to values as low as  $0.01\,\msun$ per AGB star and $0.02\,\msun$ per SN \citep[Fig.~1 of][]{michalowski15}. This means that both populations might in principle be responsible for dust production. However, for star-forming galaxies at $z\sim6.34$ and $7.5$ the required yields per AGB stars are as high as $0.1$--$1\,\msun$, and per SN above $1\,\msun$. This implies that another mechanism  must be responsible for dust production in these galaxies.

Grain growth in the ISM is the most likely alternative. This process is believed to be very fast with a timescale of a few tens of million years \citep{draine90,draine09,hirashita00,zhukovska08}, so it can be invoked even for galaxies observed a few hundred million years after the Big Bang.

\section{Conclusions}

I presented a method which is based on the comparison of the measured dust and stellar masses (the latter giving an estimate of the number of  dust-producing stars) in order to assess the dust origin of high redshift galaxies. This was used to show that for most of the galaxies with detected dust emission between $z=4$ and $z=7.5$ ($1.5$--$0.7$ billion years after the Big Bang) AGB stars are not numerous and efficient enough to be responsible for the measured dust masses. Supernovae could account for most of the dust, but only if all of them had efficiencies close to the maximal theoretically allowed value. This suggests that a different mechanism is responsible for dust production at high redshifts, and the most likely possibility is the grain growth in the interstellar medium.

\acknowledgements{I am grateful to the Polish Astronomical Society for the Young Scientist Award. I acknowledge the support of the UK Science and Technology Facilities Council (STFC).
This research has made use of  
the NASA's Astrophysics Data System Bibliographic Services;
and the  Edward Wright Cosmology Calculator \citep{wrightcalc}:
\tt{www.astro.ucla.edu/ $\sim$wright/CosmoCalc.html}.
}

%\input{/Users/michal/bibtex/getbibaa_like_apj.tex}

%\bibliographystyle{ptapap}
%\bibliography{ptapapdoc}

\end{document}